\documentstyle[epsfig,pra,aps,tighten,twocolumn]{revtex}

\begin{document}
\draft
\title{\bf Steady state behaviour in atomic three-level lambda and
ladder systems with incoherent population pumping\\}

\author{M. Blaauboer} 

\address{Faculteit Natuurkunde en Sterrenkunde, Vrije Universiteit,
         De Boelelaan 1081, 1081 HV Amsterdam, The Netherlands}

\date{\today}
\maketitle

\begin{abstract}
\normalsize{We study the steady state in three-level 
lambda and ladder systems. It is well-known that in a lambda system 
this steady state is the coherent population trapping state, 
independent of the presence of spontaneous emission. In contrast, 
the steady state in a ladder system is in general not stable against 
radiative decay and exhibits a minimum in the population 
of the ground state. We show that incoherent population pumping
destroys the stability of the coherent population trapping state in 
the lambda system and suppresses a previously discovered sharp dip 
in the steady state response. In the ladder system the observed 
minimum disappears in the presence of an incoherent pump on 
the upper transition.}
\end{abstract}

\pacs{PACS number(s): 42.50.Hz
       {\tt nlin-sys/9701002}} 
\narrowtext 

\begin{figure}
\centerline{ \epsfig{figure=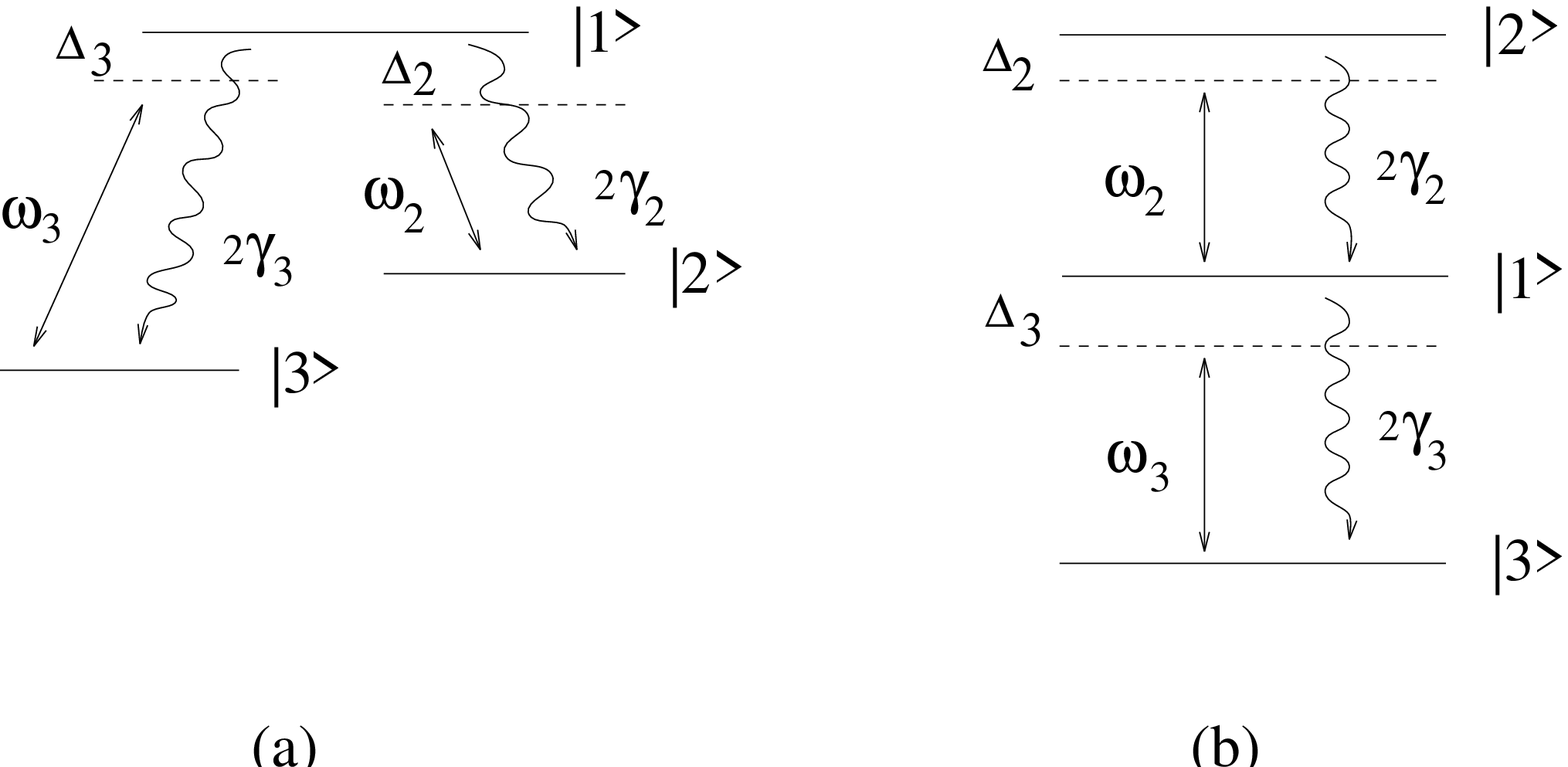,width=0.95\hsize}}
\caption{
Schematic representation of a three-level $\Lambda$ (a) and 
three-level ladder (b) atomic system.}
\label{fig:systems}
\end{figure}

%
The first observation in an optical pumping experiment of what is 
now known as the phenomenon 
of coherent population trapping (CPT) was made in 1976 
by Alzetta {\it et al.} \cite{alzetta76}. They found that the fluorescence
intensity of sodium vapour, illuminated by a laser
and analyzed as a function of an applied magnetic
field, decreased when the difference in frequency of two laser modes matched
some hyperfine transitions of the ground state 
of sodium. A
theoretical explanation for this observation can be given in terms
of a three-level atomic system in which coherent radiation fields
couple two ground states, the initial and final state, to a common 
excited state. Under the two-photon Raman resonance
condition, when the frequency difference of the radiation fields equals 
the separation between the initial and final level,
population becomes trapped in a coherent superposition of these 
two levels. Since its discovery, 
many studies of CPT have appeared using three-level lambda ($\Lambda$), 
ladder and $V$ systems\cite{orriols79,radmore82,dalton82,scully89,boublil91}.
In addition, the concept of CPT has found interesting applications
in, for example, the fields of laser cooling\cite{aspect88}, adiabatic 
population transfer\cite{kuklinski89} and lasing without
inversion\cite{imamoglu91,agarwal93,vemuri94,vemuri96}. \\
Recently, Jyotsna and Agarwal\cite{jyotsna95} studied the dynamics of the 
populations in a three-level $\Lambda$ system with spontaneous emission 
(cf. FIG.~\ref{fig:systems}(a)). They demonstrated that 
the steady state reached by the system is always the CPT state, 
irrespective of the strengths of the fields and the rates of
spontaneous emission. For weaker fields it only takes longer
times to reach the CPT state.
Motivated by this result, we investigate here
the influence of incoherent population pumping on
one of the two transitions in the $\Lambda$ system.
It will be shown that in the presence 
of this additional pumping mechanism, the steady state  
is no longer the CPT state and that a sufficiently strong rate of
incoherent pumping  suppresses the dip which was found in the steady state
response of a $\Lambda$ system as a function of detuning from the two-photon
resonance condition\cite{jyotsna95}. Besides, we show that incoherent 
population pumping has an interesting effect
on the steady state population
in another three-level configuration, the ladder system 
(FIG.~\ref{fig:systems}(b)). It is well-known that in the 
absence of radiative decay the ladder system also gives 
rise to coherent population trapping\cite{radmore82}. 
Spontaneous emission from the upper level destroys this
trapping behaviour and leads to a minimum in the population
of the lower level. We demonstrate that the minimum disappears 
in the presence of an incoherent pump.

\begin{figure}
\centerline{\epsfig{figure=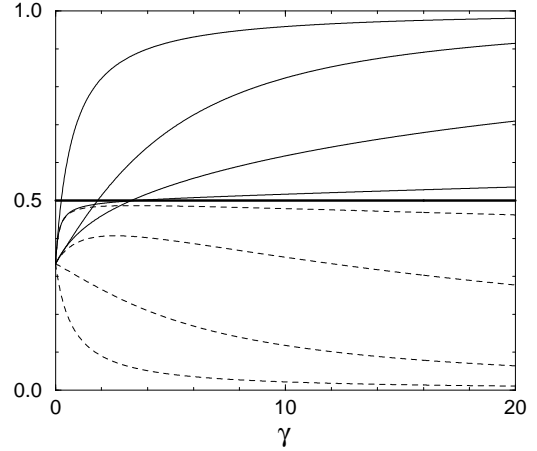,width=0.95\hsize}}
\caption{ 
$\rho_{22}$ (dashed curves) and $\rho_{33}$ (solid curves) in the 
$\Lambda$ configuration as a 
function of $\gamma_{2} = \gamma_{3} \equiv \gamma$.
From top to bottom (bottom to top) the solid (dashed) curves correspond
to $\lambda=100,10,1$ and $0.1$ respectively. For $\lambda=0$, both populations
are equal to 1/2 and independent of $\gamma$, which is indicated by
the thick line in the middle of the figure. Parameters used are
$\Omega_{2} = \Omega_{3} =10$ and $\Delta_{2} = \Delta_{3} =0$.
}
\label{fig:lam5}
\end{figure}
\begin{figure}
\centerline{\epsfig{figure=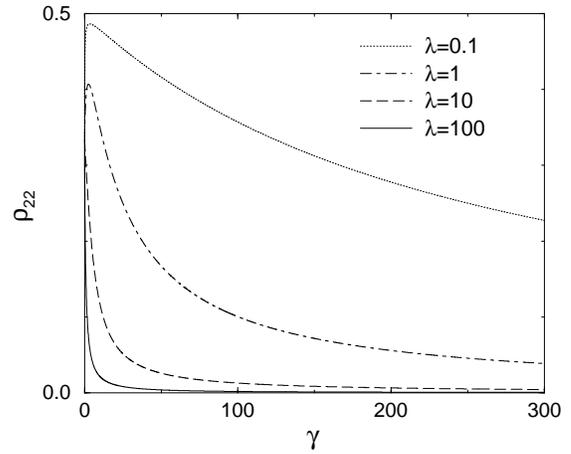,width=0.95\hsize}}
\caption{ 
$\rho_{22}$ in the $\Lambda$ system as a function of $\gamma_{2} = \gamma_{3} 
\equiv \gamma$. Parameters used are $\Omega_{2} = \Omega_{3} = 10$ and 
$\Delta_{2} = \Delta_{3} =0$.
}
\label{fig:lam6}
\end{figure}

Consider the three-level $\Lambda$ system schematically depicted in 
FIG.~\ref{fig:systems}(a).
Two states $|2\rangle$ and $|3\rangle$ are coupled to state $|1\rangle$ 
via classical monochromatic light fields of frequencies
$\omega_{2}$ and $\omega_{3}$ respectively. Level $|1\rangle$ 
decays to the lower-lying level $|2\rangle$ ($|3\rangle$)
by spontaneous emission with rate $2\gamma_{2}$ ($2\gamma_{3}$).
Let $\Omega_{2}$ and $\Omega_{3}$, defined as $\Omega_{j}=\mu_{1j} 
{\cal E}_{j} / \hbar$ ($j=2,3$), denote the Rabi frequencies associated 
with the coupling by the fields, with $\mu_{1j}$ the dipole matrix
interaction term between level 1 and level $j$, and ${\cal E}_{j}$ 
the amplitude of laser field $j$. 
The Hamiltonian for this system in the appropriate rotating frame
can be written as
\begin{eqnarray}
{\cal H} & = & \hbar \Delta_{3} |1\rangle \langle 1| + \hbar 
(\Delta_{3} - \Delta_{2}) |2\rangle \langle 2| + \nonumber \\
& & - \left( \frac{1}{2} \hbar \Omega_{2} |1\rangle \langle2| +
\frac{1}{2} \hbar \Omega_{3} |1\rangle \langle 3| + \mbox{\rm H.c.} \right),
\end{eqnarray}
with $\Delta_{2} = E_{1} - E_{2} - \hbar \omega_{2}$ and 
$\Delta_{3} = E_{1} - \hbar \omega_{3}$.
We have chosen the energy of level $|3\rangle$, $E_{3}$, as the zero of energy.
The density matrix equations are obtained
from the Liouville equation including damping effects\cite{agarwal74}
and have partly been described elsewhere\cite{imamoglu91,vemuri94,jyotsna95}.
In the presence of an incoherent pump\cite{incoherent}
of rate $\lambda$ on the $|1\rangle$-$|2\rangle$ transition
they are given by
\begin{mathletters}
\begin{eqnarray}
\dot{\rho}_{11} & = & - 2 (\gamma_{2} + \gamma_{3}) \rho_{11} 
- \left( \frac{i}{2} \Omega_{2}^{*} \rho_{12} + \frac{i}{2} 
\Omega_{3}^{*} \rho_{13} + \mbox{\rm H.c.} \right) + \nonumber \\
& &
- 2 \lambda (\rho_{11} - \rho_{22}) \\
\dot{\rho}_{12} & = & - \left( \gamma_{2} + \gamma_{3} - i\Delta_{2} \right) 
\rho_{12} + \frac{i}{2} \Omega_{2} (\rho_{22} - \rho_{11}) + \nonumber \\
& &
\frac{i}{2} \Omega_{3} \rho_{32} - 2\lambda \rho_{12} \\
\dot{\rho}_{13} & = & - \left( \gamma_{2} + \gamma_{3} - i\Delta_{3} \right) 
\rho_{13} + \frac{i}{2} \Omega_{2} \rho_{23}  + \nonumber \\
& &
\frac{i}{2} \Omega_{3} (1 - 2\rho_{11} - \rho_{22}) - \lambda \rho_{13} \\
\dot{\rho}_{22} & = &  2 \gamma_{2} \rho_{11}  
+ \left( \frac{i}{2} \Omega_{2}^{*} \rho_{12} + \mbox{\rm H.c.} \right) + 
2 \lambda (\rho_{11} - \rho_{22}) \\
\dot{\rho}_{23} & = &  i (\Delta_{3} - \Delta_{2}) 
\rho_{23} + \frac{i}{2} \Omega_{2}^{*} \rho_{13} - \frac{i}{2}
\Omega_{3} \rho_{21} - \lambda \rho_{23}
\end{eqnarray}
\label{lamalg}
\end{mathletters}
For $\Delta_{2} = \Delta_{3} = 0$ and $\Omega_{2} = \Omega_{3} \equiv \Omega$
real, this leads to the steady state solution
\begin{mathletters}
\begin{eqnarray}
\rho_{11} & = & \frac{\Omega^2 \lambda}{N} \\
\rho_{22} & = & \frac{\Omega^2 T}{2 N M} \\
\rho_{33} & = & 1 - \rho_{11} - \rho_{22} \\
\mbox{\rm Im}(\rho_{12}) & = & \frac{2}{\Omega} ( (\gamma_{2} + \lambda) 
\rho_{11} - \lambda \rho_{22}) \\
\mbox{\rm Im}(\rho_{13}) & = & \frac{2\gamma_{3}}{\Omega} \rho_{11} \\
\mbox{\rm Re}(\rho_{23}) & = &  \rho_{22} - \frac{(\gamma_{2} + \gamma_{3} + \lambda)}
{\lambda}\rho_{11} \\
\mbox{\rm Re}(\rho_{12}) & = & \mbox{\rm Re}(\rho_{13}) = 
\mbox{\rm Im}(\rho_{23}) = 0
\end{eqnarray}
\label{steadylam}
\end{mathletters}
where
\begin{eqnarray}
M & = & 2 \lambda (\gamma_{2} + \gamma_{3} + 2 \lambda) + \Omega^2 \nonumber \\
N & = & (4\gamma_{3}\lambda + \Omega^2)(\gamma_{2} + \gamma_{3} + \lambda) + 
2\lambda \Omega^2 \nonumber \\
T & = & (\gamma_{2} + \gamma_{3} + 2\lambda)(4 \lambda(\gamma_{2} + \lambda) 
+ \Omega^2). \nonumber
\end{eqnarray}
In the absence of the incoherent pump, so $\lambda=0$, the steady state
(\ref{steadylam}) equals the CPT state, which has been studied 
by Jyotsna and Agarwal\cite{jyotsna95}. In the $\Lambda$ system with 
symmetric fields the population is then equally distributed 
between the two ground levels. Switching on the incoherent pump and
taking equal rates of spontaneous emission $\gamma_{2}=\gamma_{3}\equiv
\gamma$, we have plotted in FIG.~\ref{fig:lam5} the 
steady state populations $\rho_{22}$ (dashed curves) and $\rho_{33}$
(solid curves) as a function of the rate of spontaneous emission
for various pumping rates.
As soon as $\lambda\neq 0$, the distribution of population over the two
ground states becomes asymmetric ($\rho_{33}>\rho_{22}$) and the higher $\lambda$,
the larger the difference between the two. In fact, as $\gamma \rightarrow 
\infty$, all curves for $\rho_{33}$ go to 1, and those for $\rho_{22}$
go to 0, except if $\lambda=0$. For sufficiently strong decay rates,
the incoherent pump thus depletes level $|2\rangle$.
This can be seen more clearly  in 
FIG.~\ref{fig:lam6}, which shows $\rho_{22}$ on a larger scale for $\gamma$.
From (\ref{steadylam}) we obtain in the limit 
$\gamma \lambda \gg \Omega^2$ that
\begin{eqnarray}
\rho_{11} & \simeq & \frac{\Omega^2}{4\gamma (2\gamma + \lambda)} \simeq 0 \nonumber \\
\rho_{22} & \simeq & \frac{\Omega^2 (\gamma + \lambda)}{4\gamma \lambda (2\gamma + \lambda)} 
\simeq 0 \nonumber \\
\rho_{33} & \simeq & \frac{4\gamma \lambda (2\gamma + \lambda) - \Omega^2 
(\gamma + 2\lambda)}{4\gamma \lambda (2\gamma + \lambda)} 
\simeq 1. \nonumber 
\end{eqnarray}
FIG.~\ref{fig:lam5} also shows that 
$\rho_{11}=\rho_{22}=\rho_{33}=1/3$
at $\gamma=0$, as long as $\lambda \neq 0$. In the absence of decay,
the incoherent pump distributes the population equally among the 
levels of the $\Lambda$ system, independent of $\lambda$ and $\Omega$. 

\begin{figure}
\centerline{\epsfig{figure=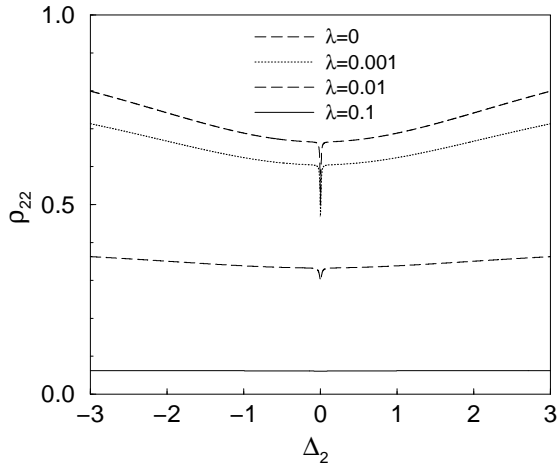,width=0.95\hsize}}
\caption{ 
The population of level $|2\rangle$ in the $\Lambda$ system as a function of 
detuning $\Delta_{2}$ for various incoherent pumping rates $\lambda$. 
The parameters used are $\Omega_{2} = \Omega_{3} = 0.2$, $\gamma_{2}=2.0$, 
$\gamma_{3}=1.0$ and $\Delta_{3}=0$.
}
\label{fig:lam7}
\end{figure}

Armed with the knowledge that incoherent population pumping leads to 
a steady state which is different from the CPT state in the $\Lambda$ 
system, one could ask
how it would affect the dependence of $\rho_{22}$ on the detuning 
$\Delta_{2}$. The question arises because Jyotsna and Agarwal\cite{jyotsna95}
found that for weak fields and unequal decay rates $\gamma_{2}$ and $\gamma_{3}$, 
$\rho_{22}$ 
exhibits a sharp dip as a function of $\Delta_{2}$ around the Raman 
resonance condition $\Delta_{2} = \Delta_{3}$. They explain that this happens 
because for unequal decay rates (say $\gamma_{2} > \gamma_{3}$), more population
is present in state $|2\rangle$ than in state $|3\rangle$ if $\Delta_{2} \neq 
\Delta_{3}$.
But at the resonance condition, state $|2\rangle$ always contains 
half of the population. Hence a dip occurs around this value, which becomes 
sharper with increasing $\gamma_{2}$. For $\lambda\neq 0$, however, 
the steady state at $\Delta_{2} = \Delta_{3}$ is no longer the CPT state 
and level $|2\rangle$ contains less than half of the total population. We have 
calculated the steady state response of the $\Lambda$ system with incoherent 
pumping for $\Delta_{2}\neq \Delta_{3}$.
The resulting expressions are lengthy and therefore not given here. However, 
the dependence of $\rho_{22}$ on $\Delta_{2}$ for weak fields 
$\Omega \ll \gamma_{2}, \gamma_{3}$ is shown in FIG.~\ref{fig:lam7}.
We see an overall decrease in $\rho_{22}$ as $\lambda$ increases. The
dip in $\rho_{22}$ completely disappears for $\lambda \simeq \Omega$ and
$\rho_{22}$ then becomes independent of the detuning. This is expected if 
incoherent pumping is the dominant pumping mechanism, since the amount of 
detuning from the transition is irrelevant for a pump with a linewidth
much larger than the transition width. The same behaviour occurs 
for stronger fields, for which the dip is smoother\cite{jyotsna95}.

\begin{figure}
\centerline{\epsfig{figure=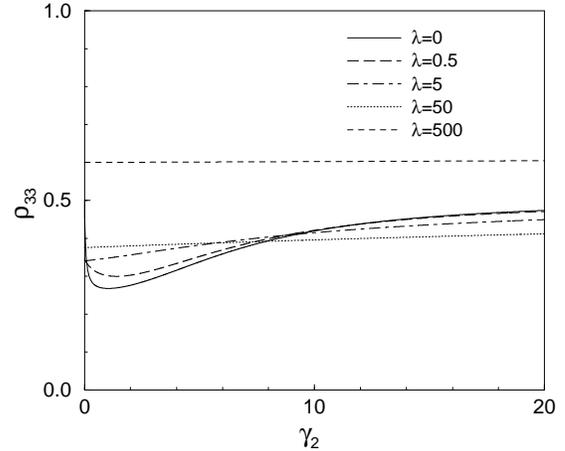,width=0.95\hsize}}
\caption{ 
$\rho_{33}$ vs. $\gamma_{2}$ in the ladder system for increasing incoherent 
pumping rates
on the $|1\rangle-|2\rangle$ transition. The parameters used are
$\Omega_{2}=\Omega_{3}=10$, $\Delta_{2} = \Delta_{3}=0$ and $\gamma_{3}=0.1$.
}
\label{fig:lad4}
\end{figure}

Let us now consider the ladder system of FIG.~\ref{fig:systems}(b) with 
an incoherent pump on the $|1\rangle$-$|2\rangle$ transition. 
The Hamiltonian and the evolution equations of the reduced density matrix 
for this system are similar to those of the $\Lambda$ configuration and 
not given here (see eg. \cite{agarwal93,vemuri94}). From them the steady 
state solution is easily obtained. FIG.~\ref{fig:lad4} shows 
the steady state population of level $|3\rangle$ as a function of 
$\gamma_{2}$ for symmetric fields and under the Raman resonance condition.
We see that $\rho_{33}$ exhibits a minimum in the absence of
incoherent population pumping.
This minimum arises because as soon as $\gamma_{2} \neq 0$, the steady
state in the ladder system deviates from the CPT state and level
$|1\rangle$ becomes populated. For small values of 
$\gamma_{3}$\cite{condition}, level $|3\rangle$ 
contributes to the population in level $|1\rangle$ and so $\rho_{33}$ initially
decreases with $\gamma_{2}$. However, as $\gamma_{2}$ increases further,
$\rho_{33}$ approaches again the value 1/2
and hence the minimum is formed.

The size of the minimum becomes smaller with increasing $\lambda$ and 
disappears for $\gamma_{3} \ll \lambda \leq \Omega$. If the rate of 
incoherent pumping dominates the spontaneous emission from
level $|2\rangle$, one can show that
\begin{eqnarray}
\rho_{11} & \simeq & \frac{\Omega^2}{4\lambda \gamma_{3} + 3\Omega^2}  
\nonumber \\
\rho_{22} & \simeq & \frac{\Omega^2}{4\lambda \gamma_{3} + 3\Omega^2}  
\nonumber \\
\rho_{33} & \simeq & \frac{4\lambda \gamma_{3} + \Omega^2}{4\lambda \gamma_{3}
+ 3\Omega^2}. \nonumber
\end{eqnarray}
The steady state populations have become independent of $\gamma_{2}$ in 
this limit and the upper two levels contain equal amounts of population,
as expected when the coupling between these levels is large compared to
the decay. 
%

Summarizing, we analyzed the behaviour of the steady state in 
atomic three-level $\Lambda$ and ladder systems in the presence of 
incoherent population pumping. In the $\Lambda$ system the steady state 
is no longer equal to the CPT state under influence of this pump. A similar
decay of the trapping state has recently been shown to occur due to 
fluctuations between two driving fields in a double-$\Lambda$ 
configuration\cite{fleischhauer94}. We have demonstrated that
a sufficiently strong rate of incoherent pumping suppresses
the sharp dip which was found in the steady state response of
the $\Lambda$ system as a function of detuning $\Delta_{2}$ \cite{jyotsna95}.
In the ladder configuration,
the minimum which occurs in the steady state population of the lowest
level $|3\rangle$
as a function of the rate of spontaneous emission from the upper 
level $|2\rangle$ disappears if the $|1\rangle-|2\rangle$ transition 
is incoherently pumped.

Other works on three-level atomic level schemes have demonstrated
that phase-diffusion in the pumping fields has
substantial effects on coherence phenomena such as 
the gain in a lasing without inversion (LWI) ladder
system\cite{vemuri95} and the refractive index enhancement
in a $V$ configuration\cite{gong95}. These phase fluctuations in the 
driving fields are also known to lead to a decay of the CPT state
in $\Lambda$ systems\cite{dalton82,fleischhauer94} and thus their
effect seems to be qualitatively similar to that of incoherent population
pumping.
Another point of consideration is that we have 
only studied dilute media of $\Lambda$ and ladder systems here. 
In order to treat dense media, local-field corrections have to be taken into account. 
It has been shown for the coherently pumped $\Lambda$ system that 
CPT persists in dense media 
for all field strengths\cite{jyotsna96} and it would be interesting 
to see what the result is of including an incoherent pump.

\acknowledgments

I have benefited from useful discussions with G. Nienhuis.


\begin{references}
\bibitem{incoherent} A pump is named "incoherent" if its linewidth is much
                    larger than the width of the transition on which it 
                    operates\cite{vemuri94}.
\bibitem{condition} The condition for the occurrence of the minimum is
		$\gamma_{3} < \frac{1}{6} \sqrt{6} \Omega$.
\end{references}
\end{document}